\documentclass[journal=nalefd,manuscript=letter,layout=twocolumn]{achemso}

\usepackage{upgreek}
\usepackage{graphicx}
\usepackage{amssymb}
\usepackage{amsmath}
\usepackage{latexsym}
\usepackage{rotating}
\usepackage{float}
\usepackage[breaklinks=true]{hyperref}
\usepackage[labelfont=bf]{caption}
\usepackage{ragged2e}
\DeclareCaptionJustification{justified}{\justifying}
\usepackage{xcolor}

\usepackage{epstopdf}   

\hypersetup{
  colorlinks   = true,
  urlcolor     = blue,
  linkcolor    = blue,
  citecolor    = red
}

\author{Marta Cagetti}
\altaffiliation{M.C. and S.F. contributed equally to this work}
\affiliation{ICFO-Institut de Ci\`encies Fot\`oniques, The Barcelona Institute of Science and Technology, Castelldefels (Barcelona) 08860, Spain}

\author{Stefan Forstner}
\altaffiliation{M.C. and S.F. contributed equally to this work}
\affiliation{ICFO-Institut de Ci\`encies Fot\`oniques, The Barcelona Institute of Science and Technology, Castelldefels (Barcelona) 08860, Spain}
\email{stefan.forstner@icfo.eu}

\author{Victor Champain}
\affiliation{ICFO-Institut de Ci\`encies Fot\`oniques, The Barcelona Institute of Science and Technology, Castelldefels (Barcelona) 08860, Spain}

\author{Roger Tormo-Queralt}
\affiliation{ICFO-Institut de Ci\`encies Fot\`oniques, The Barcelona Institute of Science and Technology, Castelldefels (Barcelona) 08860, Spain}

\author{Christoffer B. M{\o}ller}
\affiliation{ICFO-Institut de Ci\`encies Fot\`oniques, The Barcelona Institute of Science and Technology, Castelldefels (Barcelona) 08860, Spain}
\alsoaffiliation{Present address: Niels Bohr Institute, University of Copenhagen, 2100 Copenhagen, Denmark}

\author{Sergio L. De Bonis}
\affiliation{ICFO-Institut de Ci\`encies Fot\`oniques, The Barcelona Institute of Science and Technology, Castelldefels (Barcelona) 08860, Spain}

\author{Chandan Samanta}
\affiliation{ICFO-Institut de Ci\`encies Fot\`oniques, The Barcelona Institute of Science and Technology, Castelldefels (Barcelona) 08860, Spain}
\alsoaffiliation{Present address: Department of Physics, Indian Institute of Science Education and Research Bhopal (IISER Bhopal), Bhopal 462066, India}

\author{Elsa V\'azquez-Rodriguez}
\affiliation{ICFO-Institut de Ci\`encies Fot\`oniques, The Barcelona Institute of Science and Technology, Castelldefels (Barcelona) 08860, Spain}

\author{Eneko Mateos-Madinabeitia}
\affiliation{ICFO-Institut de Ci\`encies Fot\`oniques, The Barcelona Institute of Science and Technology, Castelldefels (Barcelona) 08860, Spain}

\author{David A. Czaplewski}
\affiliation{Center for Nanoscale Materials, Argonne National Laboratory, Argonne, Illinois 60439, United States}

\author{Adrian Bachtold}
\affiliation{ICFO-Institut de Ci\`encies Fot\`oniques, The Barcelona Institute of Science and Technology, Castelldefels (Barcelona) 08860, Spain}
\email{adrian.bachtold@icfo.eu}

\title{Current-based RF charge sensing in a carbon nanotube}

\abbreviations{RF,DQD,ICT,HEMT,SNR,RLC}

\begin{document}

\begin{abstract}
Ultra-sensitive charge detection is a widely used tool for quantum electronics with applications
in quantum information processing and in probing the physics of condensed matter systems.
Existing approaches require either an impedance-matched resonant circuit, or millimeter-scale proximity between amplifier and sample, both adding complexity and constraining device
design. In this work, we introduce a current-mode charge sensor in a suspended carbon nanotube, operating at the $1.25~\rm MHz$ resonance of an RLC tank circuit and achieving a charge sensitivity of $0.15$~$\mu$e$/\sqrt{{\rm Hz}}$. We utilize it
to measure a double quantum dot (DQD) electrostatically defined in the same nanotube, revealing
a highly regular charge stability diagram.
We perform single-shot readout of the DQD charge state at
an integration time of $3.56$~$\mu$s, without any false assignments over $10^7$ measurements and a signal-to-noise ratio of 17 exceeding the state of the art.
\end{abstract}

\maketitle
\vspace{1ex}\hrule\newpage

Sensitive electrometers are a key requirement for quantum information processing in semiconductors \cite{Burkard2023_spinqubits_RMP,Hanson2007_spins_RMP,LossDiVincenzo1998_computation_PRA}.
Perhaps the most impactful use case is the readout of spin qubits, which is usually based on spin-to-charge conversion followed by charge state readout \cite{Elzerman2004_singleshot_Nature, Hanson2007_spins_RMP}.
Quantum error correction requires qubit readout fidelities above the relevant fault-tolerance threshold \cite{Fowler2012_surface_PRA}.
A further area of interest is probing new physics of mesoscopic and strongly correlated systems \cite{Yoo1997ScanningSET_Science}.
Charge sensing has been employed in a wide range of fundamental studies including imaging of edge states in quantum hall physics \cite{Ilani2004_microscopic_Nature}, fractional Chern insulators \cite{Xie2021FractionalChern}, and phase transitions \cite{Honig2013LocalElectrostatic_NMat,Shapir2019ImagingWignerCrystal_Nature}.

Radio-frequency reflectometry is the state-of-the-art approach to rapid, high-fidelity charge detection. In a radio-frequency single-electron-transistor (RF-SET), first introduced by Schoelkopf et al. in 1998 \cite{Schoelkopf1998_RFSET_Science}, the charge sensor is impedance matched to a 50 $\Omega$ transmission line via a resonant circuit and probed by a microwave tone. Changes in the measured system's charge configuration modify the sensor's conductance and hence the reflected microwave power \cite{Schoelkopf1998_RFSET_Science,Devoret2000SET_Nature,Reilly2007_QPC_APL}.
However, achieving high charge sensitivity and readout fidelity with such impedance-matched sensors requires demanding circuit and device engineering~\cite{Vigneau2023_RFReflectometry_APR,muller_situ_2010,apostolidis_quantum_2024}.
Dispersive gate-based sensing is an alternative to charge sensing that probes the quantum or tunneling capacitance but requires precise impedance matching~\cite{Hasler2015,Ares2016,GonzalezZalba2015_limits_NComms} and yields a signal only near charge degeneracy \cite{Colless2013_GateSensors_PRL,Ahmed2018_RadioFrequency_PRL,Ibberson2021_CMOS_PRX,Petersson2010_DQD_NanoLett,Mizuta2017_QuantumTunneling_PRB,Ibberson2019_RFResonator_APL,Niegemann_Parity_PRXQ_2022,Zheng_gate_based_NNano_2019,Delbecq_Cavity_PRL_2011,Chorley_CNT_PRL_2012,ares_resonant_2016}.

Direct current measurement of the charge sensor in the low-megahertz range provides an alternative method by amplifying the sensor current directly with a cryogenic high-electron-mobility transistor (HEMT), eliminating the need for impedance matching.
The scheme has been deployed for single-shot readout of spin qubits using charge sensors \cite{Tracy2016_APL_RCHEMTCS,Vink2007_APL_RCHEMTCS,Blumoff2022_PRXQ_RCHEMTCS,Mills2022_PRApp_RCHEMTCS}.
The principal limitations of the scheme are firstly $1/f$ noise which limits the signal-to-noise ratio (SNR), and secondly the requirement to place the amplifier within millimeters of the sample to minimize parasitic capacitance, which hinders integration and putting additional heat load on the cryostat's mixing chamber plate.

Here, we present a suspended carbon nanotube (CNT) charge sensor based on RF current measurement without impedance matching and use it to read out the charge states of a double quantum dot (DQD) defined in the same nanotube.
Rather than placing the HEMT amplifier close to the charge sensor to minimize the parasitic capacitance, we exploit this capacitance by adding an inductor to form a MHz-frequency RLC resonator.
The RLC resonator is formed by an RF transmission line and standard SMD components.
Compared to previous schemes \cite{Tracy2016_APL_RCHEMTCS,Vink2007_APL_RCHEMTCS,Blumoff2022_PRXQ_RCHEMTCS,Mills2022_PRApp_RCHEMTCS,Bohuslavskyi2022_APL_RCHEMTCS}, the operation on resonance allows us to perform
current-to-voltage conversion at resonance, so that
only noise in a frequency band around resonance contributes -- suppressing $1/f$-noise and other low-frequency noise contributions.
Moreover, our scheme allows the HEMT amplifier to be separated from the sample and placed at the 3.2 K stage rather than at the mixing chamber, simplifying the setup and reducing the heat load on the mixing-chamber plate.
The scheme is adapted from existing methods to measure the minute signal of electronic shot noise \cite{Jezouin2013_LC_HEMT_science,Bocquillon2013_LC_HEMT_science,Jullien2014_R_LC_HEMT_Nature} and motion of nanomechanical oscillators \cite{deBonis2018_Ultasensitive_Nanolett,Urgell_Cooling_NPhys_2020,samanta_nonlin_NPhys_2023}.

With this simple setup, we surpass the state of the art in charge sensitivity \cite{Ahmed2018_RadioFrequency_PRL,Brenning2006_ultrasensitive_JAP,Schaal2020_Silicon_PRL}.
Further, we use the sensor to map the charge stability diagram of a nearby double quantum dot, defined in the same carbon nanotube, and to characterize the interdot charge transition (ICT).
We then perform single-shot charge state readout with an integration time of 3.56~$\mu$s.
We record no false assignments over $10^7$ measurements
and achieve a signal-to-noise ratio (SNR) of 17 surpassing the current state-of-the-art at comparable integration time \cite{Connors2020_HighFidelityReadout_PRApp, Ibberson2021_CMOS_PRX, Vink2007_APL_RCHEMTCS,Tracy2016_APL_RCHEMTCS,Blumoff2022_PRXQ_RCHEMTCS,Huang2021}.
The results presented in this work were achieved using two different devices. We benchmark the readout quality with device A, while the interdot characterization and the stability diagram were obtained with device B. The characteristics of the devices are reported in Section S1 of the Supporting Information.

\begin{figure*}[t]
\centering
    \includegraphics[width=\textwidth]{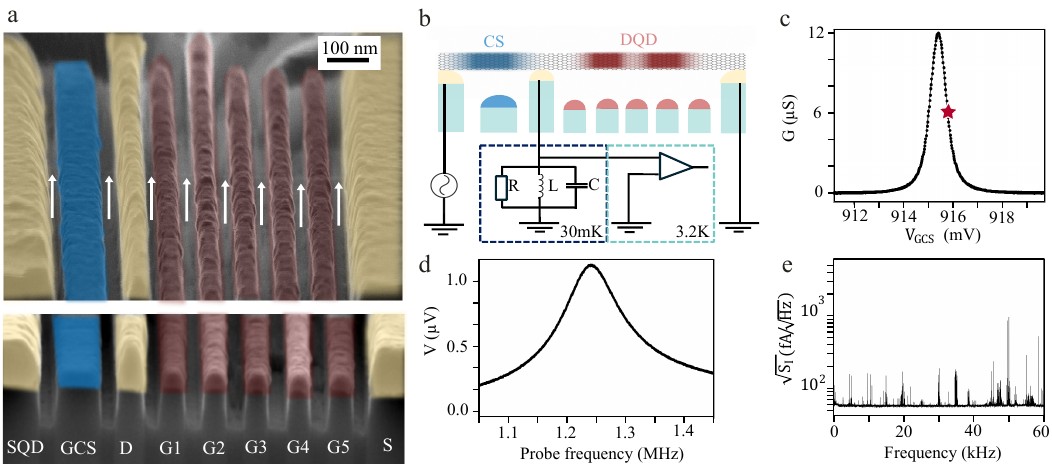}
    \caption{
(a) False-colored scanning electron micrograph of the device. The CNT, which is suspended over the gate electrodes GCS and G1-G5, is electrically in contact with sensor's source (SQD), the DQD source (S), and a common drain (D). The white arrows indicate the position of the CNT.
(b) Schematic of the setup. The common drain is connected to ground through an RLC resonator with resonance frequency $f_{{RLC}}=1.25$ MHz. The signal is amplified by a cryogenic HEMT amplifier at 3.2 K.
(c) Conductance of the sensor dot as a function of $V_\mathrm{GCS}{}$, exhibiting a Coulomb blockade peak. The red star indicates the working point for optimal sensing.
(d) Frequency response of the RLC resonator as a function of probe frequency showing a bandwidth of $\Delta f=90$~kHz.
(e) Current noise spectral density referred to the input of the voltage amplifier, obtained by demodulating the RLC output at the resonance frequency $f_{{RLC}}=1.25$ MHz, while applying an AC probe tone to the sensor source.
}
    \label{Fig:1}
\end{figure*}

A false-colored scanning electron micrograph of the suspended CNT and the underlying gates is shown in Fig.~\ref{Fig:1}a.
%
The charge sensor quantum dot is defined electrostatically by the control gate GCS, while five additional gates G1-5 control a nearby DQD.
We follow the fabrication methodology described in Ref.~\cite{TormoQueralt2022_NovelNanotube_NanoLett}.
Two regions (sensor and DQD) of the CNT are separated by a common drain electrode D, which is shunted to ground through the RLC circuit, as illustrated schematically in Fig.\ref{Fig:1}b. The sensor quantum dot is operated on the flank of the Coulomb blockade peak shown in Fig.\ref{Fig:1}c.
The RLC circuit is formed with two inductors connected in series (Coilcraft 1812CS-333XJLC) that provide a total inductance $L=66$~$\mu$H, and a resistor $R_{RLC}=10$~k$\Omega$. The capacitance is mainly given by the transmission line between the device and the HEMT $C_p=246$~pF. The input voltage is applied at the resonance frequency $f_{RLC}$ on electrode SQD via a 50~$\Omega$ matched RF and an attenuated coaxial line.
When the wave reaches the printed circuit board containing the device, it is reflected. Because the wavelength at $f_{RLC}$= 1.25 MHz exceeds 100~m in PTFE, the incident and reflected waves form an approximately uniform voltage antinode at the device. Its amplitude sets the oscillating voltage applied to electrode SQD.
The frequency response of the circuit is presented in Fig.~\ref{Fig:1}d, revealing
a resonance frequency of $f_{\mathrm{RLC}} = 1.25~\mathrm{MHz}$ and
a bandwidth of $\Delta f=90$~kHz~\cite{deBonis2018_Ultasensitive_Nanolett}.
At resonance, the current through the sensor is converted to a voltage across the effective circuit impedance $Z_{\mathrm{eff}}= 7.5\,\mathrm{k}\Omega$, which is determined primarily by the resistance of the RLC resonator. This voltage is read out by the HEMT, commercially available from cryoHEMT. Figure~\ref{Fig:1}e shows the current noise spectral density of the RLC output signal when the probe tone is applied at resonance.

To quantify the sensor performance, we extract the smallest detectable charge fluctuation of the sensor dot, referred as
charge sensitivity ~\cite{Vigneau2023_RFReflectometry_APR}.
This quantity is computed as
\mbox{$\delta q = e\,\sqrt{S_{I}}\,/\big(\Delta V_{\mathrm{GCS}}\,\cdot\,|dI/dV_{\mathrm{GCS}}|\big)$}
where $\sqrt{S_{I}}$ is the equivalent current noise
shown in Fig.~\ref{Fig:1}e, $\Delta V_{GCS}$ is the gate-voltage difference between successive Coulomb blockade peaks, corresponding to the addition of one electron to the sensor quantum dot, and $dI/dV_{\mathrm{GCS}}$ is the slope of the Coulomb blockade peak evaluated at the operating point.
The best charge sensitivity of $\delta q \approx 0.15~\mu e/\sqrt{\mathrm{Hz}}$  is obtained in the frequency bands where the noise floor is given by the flat background from the Johnson-Nyquist noise of the resistor.
This is the case in most parts of the frequency band from 20 to 600~kHz, and at frequencies above 600~kHz (for more details, see Supporting Information, Section S3).
This surpasses the current state of the art by a factor 6~\cite{Brenning2006_ultrasensitive_JAP,Hakkinen_CNT_nanolett_2015,aassime2001rfset,fujisawa2000chargenoise,lehnert2003lifetime,roschier2004noise,Vigneau2023_RFReflectometry_APR}.
Here, we apply the AC voltage of $V_{\mathrm{SQD}} = 165~\mathrm{\mu V}$, which is the largest AC voltage bias that does not modify the Coulomb conductance peak of the charge sensor.
%
%
The other parameters that we use are: $\Delta V_{GCS}=180~\mathrm{mV}$, measured output current noise $\sqrt{S_I}=71~\mathrm{fA}/\sqrt{\mathrm{Hz}}$, slope $\left|dI/dV_{GCS}\right|=2.7\,\mu\text{A}/\text{V}$.
We can convert this charge sensitivity to an effective energy sensitivity of $\delta E = (\delta q)^2/(2C_\Sigma)\approx 1.5\,\hslash$, with $\hslash$ the reduced Planck constant and $C_\Sigma$ the total capacitance of the sensor quantum dot \cite{Schoelkopf1998_RFSET_Science}. Here we have expressed the capacitance via the charging energy $E_\text{c}=e^2/(2C_\Sigma)=\alpha_\text{CS} \Delta V_{GCS}$, with $\alpha_\text{CS}=0.25$~eV/V the measured lever arm of the charge sensor gate (see Supporting Information, Section S1).
Typically, we operate at lower amplitudes of around $V_{SQD}\approx 10$~$\mu$V, which provides a flatter noise floor and a sensitivity of around 0.3-1.0~$\mu e/\sqrt{\mathrm{Hz}}$, sufficient for our measurements.
We note that while the measurements reported here are performed at a temperature of around 150~mK, the sensor remains functional at higher temperatures (with lower sensitivity) until the broadening of the Coulomb peak becomes comparable to the charging energy at tens of Kelvin.
A more detailed characterization of the charge sensitivity and of operation at higher temperature is provided in the Supporting Information.

\begin{figure*}[t]
\centering
    \includegraphics[width=0.75\textwidth]{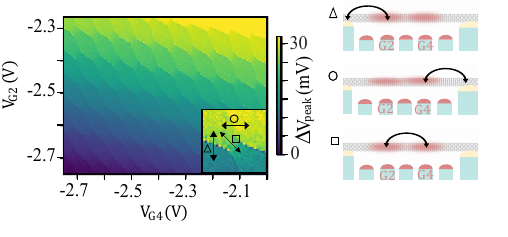}
    \vspace{-0.5cm}
    \caption{Charge stability diagram
    showing the effective position of the Coulomb peak, $\Delta V_{\mathrm{peak}}$, relative to its initial position (bottom left of the stability diagram) as a function of $V_{\mathrm{G2}}$ and $V_{\mathrm{G4}}$.
    Inset: zoom of the interdot charge transition, highlighting the three transitions, which are illustrated schematically on the right.}
    \label{Fig:2}
\end{figure*}

A suspended nanotube enables the definition of DQDs featuring a highly regular charge stability diagram \cite{TormoQueralt2022_NovelNanotube_NanoLett}. We use its small bandgap to define the three tunnel barriers required to create a DQD with gates G1, G3, and G5. We use gates G2 and G4 to tune the potential of left and right dots, respectively.
The sensor dot is capacitively coupled to the DQD, and any change in its charge configuration modifies the energy level of the sensor quantum dot, resulting in a shift of the sensor Coulomb blockade peak position.
We estimate the effective gate voltage shift $\Delta V_\text{peak}$ of the Coulomb blockade peak from the measured conductance of the charge sensor.
By tracking $\Delta V_{\rm peak}$ as a function of the G2 and G4
voltages, we reconstruct the charge stability diagram of the DQD system.
Figure~\ref{Fig:2} shows the expected highly regular honeycomb pattern measured via charge sensing, where all three possible charge transitions, schematically shown in the inset, are clearly visible.
The sensor response is stronger along $V_\mathrm{G2}$ than along $V_\mathrm{G4}$ consistent with the closer proximity of the sensor dot to the dot over gate G2.
At high charge occupation, the large applied gate voltages progressively lower the potential barrier, enhancing tunneling and smearing the ICT lines in the honeycomb pattern.

\begin{figure*}[h!]
\centering
    \includegraphics[width=0.75\textwidth]{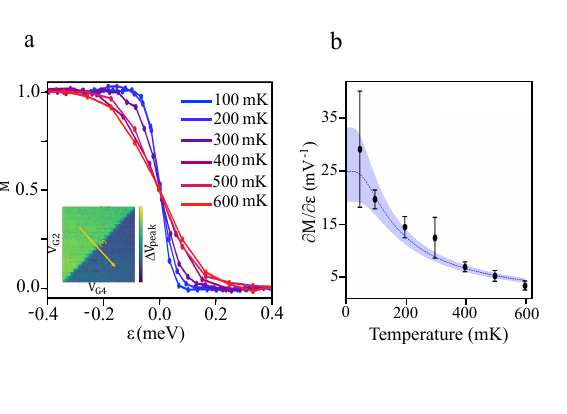}
    \vspace{-1.5cm}
\caption{
(a) Normalized average charge occupation (M) of the left quantum dot as a function of detuning for cryostat temperatures between 100 and 600~mK. The inset shows the ICT and the yellow arrow indicating the detuning axis used to extract the traces.
(b) Temperature dependence of the fitted slope parameter $\frac{\partial M}{\partial\epsilon}$, extracted from the curves shown in panel (a) at zero detuning. The dashed line is a least-squares fit to the model. The shaded region is the 95\% confidence band of the fit, computed from the covariance matrix of the fitted parameters.}
\label{Fig:3}

\end{figure*}

\begin{figure*}[h!]
\centering
    \includegraphics[width=0.75\textwidth]{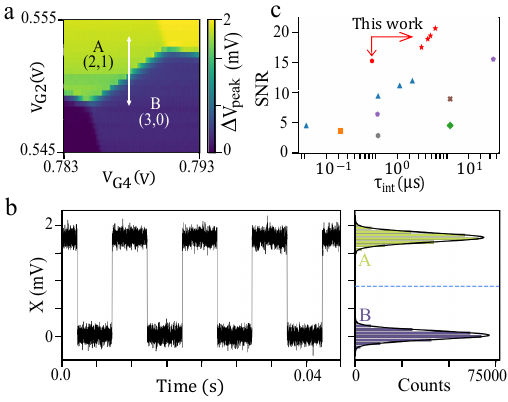}
\caption{
(a) Response of the charge sensor as a function of $V_\mathrm{G2}$ and $V_\mathrm{G4}$, showing the \((2,1)\leftrightarrow(3,0)\) ICT.
(b) Single-shot time trace of the in-phase quadrature of the demodulated lock-in output signal X, together with the corresponding histogram for an integration time of \(\tau=3.56~\mu\mathrm{s}\). The dashed line indicates the discrimination threshold.
(c) Readout SNR from this work compared with state-of-the-art charge sensors.  Red star and circle markers represent measurements performed at 1.25 MHz and 39.6 MHz, respectively. Results from previous works are shown as blue triangles (Ref.~\cite{Connors2020_HighFidelityReadout_PRApp}), an orange square (Ref.~\cite{Ibberson2021_CMOS_PRX}), grey hexagons (Ref.~\cite{Vink2007_APL_RCHEMTCS}), brown cross (Ref.~\cite{Tracy2016_APL_RCHEMTCS}), violet pentagon (Ref.~\cite{Blumoff2022_PRXQ_RCHEMTCS}), and green diamond (Ref.~\cite{Huang2021}).
}
\label{Fig:4}
\end{figure*}

We characterize the ICT by extracting the interdot tunnel coupling $t$ between the left and right quantum dots, following the method elaborated in Ref.~\cite{Petta2004Manipulation}.
The energy splitting of
the effective two-level system is given by
$\hbar\omega_{q} = \sqrt{\epsilon^2 + 4t^2}$, where $\epsilon$ is the potential difference between left and right quantum dot and $t$ is the interdot tunnel coupling. At zero detuning
$\epsilon = 0$ the splitting of the two-level system reduces to its minimum value $2t$, and
the occupation of the two states is set by the competition between
$t$ and the electron thermal energy $k_B T_e$: when $k_B T_e \ll t$ the charge
stays pinned in the ground state and the sensor resolves a sharp
switch between the two dots; as $T_e$ increases, thermal excitation
into the upper level smooths this transition, as seen in the
flattening of the curves around $\epsilon = 0$ in Fig.~\ref{Fig:3}a. These
curves were obtained by tracking the position of the sensor's
Coulomb blockade peak while sweeping across the ICT. The initial and
final peak positions correspond to an excess charge localized in
the left (M=1) and right (M=0) quantum dot, respectively. By fitting the
temperature dependence of the measured slope at zero detuning
(see Fig.~\ref{Fig:3}b) using
$\left.\frac{\partial M}{\partial\epsilon}\right|_{\epsilon=0} =
-\frac{1}{4t}\tanh\left(\frac{t}{k_{\mathrm B}T_{\mathrm e}}\right)$,
we extract the tunneling coupling $2t/h = 2.5 \pm 0.4~\mathrm{GHz}$, with $h$ the Planck constant.
The tunnel-coupling is a key-parameter for coherent manipulation of spin- and charge qubits. This measurement shows that it can be well-resolved with charge sensing, in the absence of transport.

To benchmark the quality of the charge readout, we focus on the interdot transition between two electronic states (2,1) and (3,0), shown in Fig.~\ref{Fig:4}a.
To characterize the quality of the
single-shot readout, we periodically
drive the electron between the two charge states via a square wave
applied to G4. We choose the amplitude of the square wave so that the detuning significantly exceeds the interdot tunnel coupling between the two states. The demodulated signal, integrated over a time
$\tau = 3.56~\mu\mathrm{s}$, follows the charge state of the electron
accordingly, as shown in Fig.~\ref{Fig:4}b. The right-hand side of the figure shows
the histogram of the time trace, revealing two well-separated Gaussian
distributions with means $\mu_0, \mu_1$ and standard deviations
$\sigma_0, \sigma_1$, from which we extract the signal-to-noise ratio
$\mathrm{SNR} = |\mu_1 - \mu_0|/\sigma$, with
$\sigma = (\sigma_0+\sigma_1)/2$. We
obtain $\mathrm{SNR} = 17$, exceeding the state-of-the-art reported at
comparable integration times, as shown in Fig.~\ref{Fig:4}c \cite{Connors2020_HighFidelityReadout_PRApp}.
This corresponds to a Gaussian-inferred infidelity
$1 - \mathcal{F} \sim 10^{-17}$~\cite{Blumoff2022_PRXQ_RCHEMTCS}. Over $10^7$ independent measurements,
not a single misclassification was observed.
We also observe a second resonance of the circuit at 39.6~MHz. A possible explanation is a $\lambda/2$-resonance of the transmission line between the device and the HEMT. This resonance allows for single-shot charge readout with a SNR of 15 with an integration time $\tau=0.85$~$\mu$s as shown in Fig.\ref{Fig:4}.c (see also Supporting Information Section S6).

The experimental results obtained in this work show that current-mode charge sensor-based readout produces comparable or larger signal than RF-reflectometry, while removing the need for impedance matching.
We note that an impedance matching can also be avoided by
employing a nonlinear resonator~\cite{Havir2025_Matching}.
In a circuit-level comparison of our technique to RF-reflectometry, elaborated in the Supporting Information, we find that our technique allows for improved sensitivity, whereas reflectometry naturally enables larger bandwidth.

In conclusion, we implemented a resonant RF-current readout in a carbon nanotube charge sensor, eliminating the need for impedance matching.
The simplicity of the scheme makes it particularly appealing for the study of complex samples like suspended carbon nanotubes, where proximal charge readout has barely been developed \cite{Shapir2019ImagingWignerCrystal_Nature,Hakkinen_CNT_nanolett_2015,Biercuk_2006_CNT_PRB,Churchill_2009_CNT_PRL,Chorley_CNT_PRL_2012,fedele_coupling_2025}.
Compared to previous current-readout schemes, operation on resonance with an RLC-circuit avoids 1/$f$ and 50~Hz-noise, enabling improved signal-to-noise ratio, and placement of the amplifier at the cryostat's 3.2~K stage, reducing heat-load on the mixing chamber.
Using a sensing dot embedded in the CNT, we employ the scheme to measure the charge stability diagram of the double quantum dot and the energy splitting of the electronic two-level system associated with an ICT.
%
The technique offers an alternative readout of the DQDs in nanotubes as compared to dispersive gate-based sensing \cite{Moller2026Nonlinear}.
We quantify the charge sensitivity to $\delta q \approx 0.15~\mu e/\sqrt{\mathrm{Hz}}$ and a signal-to-noise ratio of $17$ at $\tau=3.56$~$\mu$s, both surpassing the current state of the art.
The method's minimal requirements make it readily applicable to a wide range of semiconductor quantum dot systems.

\begin{acknowledgement}
We acknowledge support from the European Union (advanced ERC QTube 101198268 and Pathfinder MECH-QUBIT 101257982). Views and opinions expressed are however those of the author(s) only and do not necessarily reflect those of the European Union or the European Research Council Executive Agency. Neither the European Union nor the granting authority can be held responsible for them. Work performed by DAC at the Center for Nanoscale Materials, a U.S. Department of Energy Office of Science User Facility, was supported by the U.S. DOE, Office of Basic Energy Sciences, under Contract No. DE-AC02-06CH11357. SF, MC, and CBM acknowledge Marie Sklodowska-Curie grant agreements No. 101105814, No. 847517, and No. 101023289, respectively. EVR acknowledges the predoctoral program AGAUR-FI ajuts (2025 FI-2 00011) Joan Oro and EMM acknowledges the predoctoral program FI-STEP (2025 STEP 00040), which are backed by the Secretariat of Universities and Research of the Department of Research and Universities of the Generalitat of Catalonia, as well as the European Social Plus Fund. We also acknowledge support from MICINN Grant PID2021-122813OB-I00 and PID2024-158411OB-I00 funded by MICIU/ AEI / 10.13039/501100011033/ FEDER/UE, the Quantera grant (PCI2022-132951), the Fondo Europeo de Desarrollo, the Spanish Ministry of Economy and Competitiveness through Quantum CCAA, TED2021-129654B-I00, EUR2022-134050, MCIN with funding from European Union NextGenerationEU(PRTR-C17.I1), Fundacio Cellex, Fundacio Mir-Puig, Generalitat de Catalunya through CERCA, 2021 SGR 01441.
\end{acknowledgement}

\begin{suppinfo}
The Supporting Information will be made available with the published article.
\end{suppinfo}

\section*{Data availability}
The code used for data acquisition and analysis in this study is openly available on GitHub at Ref. \cite{GIT}. The data that support the findings are available upon request.

%
%

\twocolumn
\bibliography{bibliography}

\end{document}